# Millimeter Wave Scattering from Neutral and Charged Water Droplets

A. Heifetz*, H.T. Chien, S. Liao, N. Gopalsami, A.C. Raptis
*Nuclear Engineering Division, Argonne National Laboratory, 9700 South Cass Avenue, Argonne, IL 60439.*
**Corresponding author aheifetz@anl.gov, 630-252-4429, 630-252-3250 (fax)*

## Abstract
We investigated 94GHz millimeter wave (MMW) scattering from neutral and charged water mist produced in the laboratory with an ultrasonic atomizer. Diffusion charging of the mist was accomplished with a negative ion generator (NIG). We observed increased forward and backscattering of MMW from charged mist, as compared to MMW scattering from an uncharged mist. In order to interpret the experimental results, we developed a model based on classical electrodynamics theory of scattering from a dielectric sphere with diffusion-deposited mobile surface charge. In this approach, scattering and extinction cross-sections are calculated for a charged Rayleigh particle with effective dielectric constant consisting of the volume dielectric function of the neutral sphere and surface dielectric function due to the oscillation of the surface charge in the presence of applied electric field. For small droplets with (radius smaller than 100nm), this model predicts increased MMW scattering from charged mist, which is qualitatively consistent with the experimental observations. The objective of this work is to develop indirect remote sensing of radioactive gases via their charging action on atmospheric humid air.

## 1. Introduction
Electromagnetic (EM) scattering from dielectric and conducting particles has been extensively investigated [1-3]. However, scattering from charged dielectric aerosols, that frequently occurs in atmospheric physics [4] has received relatively little attention. In this paper, we investigate 94GHz millimeter wave (MMW) scattering properties of diffusion-charged water droplets. MMW spectral band is used for passive and active remote sensing because of low atmospheric extinction of EM waves in this band, and high sensitivity to materials measurements. In particular, 94GHz frequency corresponds to an atmospheric spectral window.

We recently observed a new phenomenon involving MMW reflection and transmission signal changes due to charges deposited on dielectric cylinders [5]. We conducted an experiment in which a negative ion generator (NIG) was placed between two horn antennas operating at 94GHz, and a dielectric cylinder was placed in the MMW path [5]. We confirmed that the cylinder was undergoing repeated charging and discharging cycles when the NIG was on or off, respectively. MMW observation of reflection and transmission showed consistent increase and decrease, respectively, whenever NIG was turned on or off. The mechanism of scattering has not been completely understood. Since at room and atmospheric humidity levels, surfaces of most dielectrics are covered with thin layer of water [6,7], we hypothesize that the problem may be reduced to the understanding of charged water interaction with EM waves.

In this paper we present the experimental results for 94GHz millimeter wave scattering from neutral and charged water mist produced with an ultrasonic atomizer. We show that scattering of MMW from the mist increases when the mist is irradiated with negative charges. We attempt to explain the experimental observations using a classical electrodynamics model of scattering from a dielectric sphere with diffusion-deposited mobile surface charge [8,9]. In this model, scattering and extinction cross-sections are calculated for a charged particle with the effective dielectric constant that consists of the volume dielectric function of the neutral sphere, and the surface dielectric function due to the oscillation of the surface charge. The surface dielectric function is obtained from the damped harmonic motion model of an electron. We use the double Debye model of frequency and temperature dependent permittivity of bulk water as a model of dielectric water sphere [10]. The number of charges is calculated using the model of atmospheric diffusion charging, which estimates the steady-state average number of charges on a water droplet as a linear function of the droplet radius [11]. To calculate the temperature dependent damping constant, we introduce the classical mechanics model of a linear viscous drag on a particle in water.

Potential applications of this work may include the ability to remotely detect the presence of radioactive gases that are emitted as byproducts of nuclear fuel cycle reactions. Such radionuclides have no distinct spectral lines in the EM spectrum. Atmospheric non-equilibrium cold plasma produced by radioactive decay of unstable isotopes has typical plasma frequency in the kHz band, and it is highly unstable due to recombination and diffusion [12]. On the other hand, the capacitive action of atmospheric fog water droplets to absorb and store charges, provided there is a difference in the scattering properties related to charge, may enable remote detection of radionuclides with MMW. A similar hypothesis was presented in [13], where formation of a large number of charged atmospheric water clusters following release of nuclear radiation was predicted, but the mechanism of scattering was not presented. We present in this paper 94 GHz MMW tests and a theoretical model that qualitatively explains the test results of charged microdroplets.

## 2. Experimental measurements

We examined the millimeter-wave transmission and reflection of charged and neutral mists using the setup given in Fig. 1. The millimeter wave setup consists of a 94GHz radiation transmitted and received in open air between a pair of Standard Gain horns. A Gunn diode oscillator was used to generate continuous waves and a pair of Schottky diode detectors was used to measure the transmitted/reflected power. An ultrasonic atomizer (UA) (commercial room humidifier) generated a plume of fine mist in the beam path of millimeter waves. We used de-ionized distilled water to produce a neutral mist. The UA produced a room-temperature mist, which eliminated the possibility of MMW scattering from thermal gradient. The mist was diffusion-charged using a negative ion generator (NIG) by applying a high voltage (-18 kV) in parallel at the tips of several wires pointed in the air. Tips of the NIG wires were placed in the middle of the UA generated mist. NIG is acting as a cold cathode emitting electrons, which are quickly absorbed by air molecules to form air ions [12]. The question of whether the mist is charged by ions or by electrons is not completely resolved. Since in the experimental setup the tip of the NIG was in immediate proximity to the mist, we assumed that free electrons were charging the mist.

Several experiments were conducted to test the MMW scattering properties of charged versus neutral mist. The experiments were performed for two scenarios: continuously flowing convective and diffusive mist, and stationary diffusive only mist. In the case of the flowing mist, measurements were made while the UA was continuously operating. In the case of the stationary mist, UA would be humidifying the air in the setup for several minutes, but measurements were taken immediately after the UA was turned off. The test scheme for either case consisted of monitoring the change in millimeter-wave transmitted and reflected signals, integrated over 1s intervals, for a time sequence of NIG turning on and off. We confirmed that in the dry air, MMW transmission or reflection was not changing to a measurable level when NIG was turning on or off. Next, with the flowing mist, we observed a consistently higher transmission of MMW through the mist when NIG was on (charged mist), compared to the case when the NIG was off (neutral mist). Figure 2(a) gives a plot for a typical case of the measured MMW transmitted intensity $T$ in arbitrary units (a.u.) as a function of time. Fig. 2(b) displays corresponding time readings of a charge counter placed next to the setup. The same measurements were repeated for the stationary mist case. The results for detected MMW transmitted intensity $T$ (a.u.) are plotted in Fig. 3(a), and the corresponding time readings of the charge counter are displayed in Fig. 3(b). In order to detect backscattered MMW, we placed a receiver antenna and detector, coupled to a low noise 20dB preamplifier, next to the emitter antenna (bistatic geometry). The results for detected MMW reflected intensity $R$ (a.u.) from flowing mist are plotted in Fig. 4, and reflected intensity measurements for stationary mist are plotted in Fig. 5.

It can be seen from the counter readings during the "off" cycles of NIG in Figs. 2(b), 3(b), 4(b) and 5(b) that the mist does not carry a significant amount of charge. Figures 2(a) and 3(a) indicated that relative changes in MMW forward scattering amplitude in response to mist charging for either flowing or stationary mist are approximately the same. Similarly, Figs. 4(a) and 5(a) indicate that relative changes in MMW backscattering amplitude in response to ionization are approximately the same for flowing and stationary mist cases. Compared to our previous studies of MMW scattering from charged cardboard cylinder [5], for the case of charged humid air, the amplitudes of MMW forward and backscattering scattering signal changes are smaller, but the response time is faster. Also, in [5], reflection and transmission signals were on the same order of magnitude, while in scattering from charged mist, backscattering is two orders of magnitude smaller than forward scattering. Possible reasons for that are examined in Section 3.4.

## 3. Model of millimeter wave scattering from charged water droplets

In general, electromagnetic scattering from random medium, such as mist, changes if the dielectric properties of the medium change in response to external stimulus. In one case, refractive index distribution function of the medium changes, i.e., average value of refractive index is the same but internal structure of the medium changes [14]. In another case, average refractive index of the medium changes at the molecular level, i.e., dielectric polarizability of the medium changes [14]. Both processes mentioned above can contribute to observed change in forward and backscattering of MMW from mist in response to ionization. In our paper, we chose to consider the mechanism of ionization-related change in dielectric polarizability using the hypothesis of charge-related surface dielectric function of water droplets. We will develop a model of ionization-related change in refractive index distribution function of the mist, and the effect of such change on scattering of MMW in our future work.

In order to interpret the experimental results, in this paper we investigate scattering and extinction cross-sections of charged and uncharged water droplets (extinction is a sum of absorption and scattering). Radar scattering and extinction cross-section efficiencies of sub-micron size spherical droplets for incident MMW follow the Rayleigh law [1,2]:

$$Q_{sca}(x,T) = (8/3)x^4 \,|[\varepsilon_{eff}(x,T)-1]/[\varepsilon_{eff}(x,T)+2]|^2, \quad (1)$$

$$Q_{ext}(x,T) = 4x\,\mathrm{Im}\{[\varepsilon_{eff}(x,T)-1]/[\varepsilon_{eff}(x,T)+2]\}, \quad (2)$$

where $a$ is the radius of the particle, $x=2\pi a/\lambda$ is the size parameter. Note that for a small value of the size parameter $x$, extinction is approximately equal to absorption. The effective dielectric function is

$$\varepsilon_{eff}(f,T) = \varepsilon_v(f,T) + \varepsilon_s(f,T), \quad (3)$$

where $\varepsilon_v(f,T)$ is the frequency and temperature dependent volume dielectric constant of bulk water, and $\varepsilon_s(f,T)$ is the frequency and temperature dependent surface dielectric constant. We used a well accepted phenomenological double Debye model of frequency and temperature dependent dielectric properties of water [11], which is valid in the spectral range from 1GH to 1THz and temperature range -20°C to 60°C. For the laboratory conditions of $T$=20°C and $f$ = 94GHz, this model gives the value of the dielectric constant $\varepsilon_v$ =7.69 + $i$ *13.32.

*3.1. Diffusion charging of water microdroplet*

Diffusion charging of a dielectric sphere due to an external source of charges deposits a layer of surface electrons on the sphere. As discussed in [15], electrons are confined to the single molecule-thick top-most layer of the dielectric water sphere. Existing models of aerosol diffusion charging suggest that the average number of charges may be proportional to the radius or the square of the radius of the droplet. The microscopic electrostatic model of water charging supports the linear dependence law. On the surface of a microdroplet, polar water molecules are oriented so that oxygen atoms with excess negative charge point inward, while hydrogen atoms with excess positive charge point outward [7]. Thus, the surface of a microdroplet can be considered as a collection of dipoles with the same orientation, where the potential difference between the layer of positive charge on the surface and the layer of negative charge just below the surface is $\Delta\varphi$ = 0.5 V. (The value of $\Delta\varphi$ = 0.25V is quoted in [11], however more recent results from molecular dynamics (MD) simulations suggest that $\Delta\varphi$ = 0.5V [7]). Therefore, a water microdroplet acts as a capacitor that can accumulate a net negative surface charge. In the steady state, diffusion charging deposits a net average negative charge [11]

$$Q \approx -4\pi\varepsilon_0 \Delta\varphi a, \quad (5)$$

where $\varepsilon_0$ is the free space permittivity and $a$ is the radius. Hence, the total average number of mobile surface electrons can be estimated as (with $\Delta\varphi$ = 0.5V)

$$N = Q/e \approx 2\pi\varepsilon_0 a/e, \quad (6)$$

where $e$ is the elementary charge.

*3.2 Surface dielectric constant*

To calculate the response of charged droplets, we consider classical electrodynamics-based modified Mie model of scattering from a dielectric sphere with free surface charge. Modification to conventional Mie theory is obtained via the equation of continuity of tangential magnetic fields at the boundary of the sphere [8]

$$\hat{n} \times (\vec{H}_1 - \vec{H}_2) = \vec{K}, \quad (7)$$

where $\vec{K}$ is the surface current density which can be related to the mobile surface charge as

$$\vec{K} = \sigma_s \vec{E}_t = \rho_s \vec{u}, \quad (8)$$

where $\sigma_s$ is surface conductivity, $E_t$ is the tangential component of the applied EM field at radial frequency $\omega$, $\rho_s$ is surface charge density, and $u$ is the tangential velocity of the charge carriers. The former can be calculated using the damped driven oscillator model. The equation of motion of mobile surface charge acted upon by the driving force $eE_t e^{-i\omega t}$ and velocity-dependent resistive force –*bu* is given as

$$\dot{u} + \gamma u = -(e/m_e) E_t e^{-i\omega t}, \quad (9)$$

where, $e$ is the charge of electron, $m_e$ is the mass of electron, and $\gamma = b/m_e$ is the phenomenological damping constant). Note that this model assumes continuum surface charge density, which is correct for microscopic sphere. In the nanoscale regime, however, the sphere has a discrete number of surface charges, and Coulomb electron-electron repulsion may need to be accounted for. These corrections will be introduced into the model in our future work.

To obtain the Rayleigh cross-section, the modified Mie coefficients can be expanded in power series in the size parameter, and retaining the lowest order term [8]. Then, it has been shown in [8] that the surface dielectric function is

$$\varepsilon_s = -\omega_s^2 / (\omega^2 + i\omega\gamma). \quad (10)$$

Here the surface plasma frequency is

$$\omega_s^2 = Ne^2 / 2\pi a^3 m_e \varepsilon_0, \quad (11)$$

where $N$ is the total number of mobile surface charges. Thus, $\varepsilon_s$ critically depends on the values of $N$ and $\gamma$. Using the expression for $N$ obtained in Eq. (6), we have

$$\omega_s^2 = e/m_e a^2, \quad (12)$$

so that that $\omega_s \propto 1/a$.

*3.3. Model of damping constant in equation of motion*

The model in [8] suggested approximating the temperature-dependent damping constant using an empirical expression $\gamma(T) \approx k_B T/\hbar$, where $k_B$ is the Boltzmann constant and $\hbar$ is Planck's constant. However, this model does not relate the damping constant to the properties of the medium. In this paper, we introduce a classical-mechanics model of the temperature-dependent damping constant $\gamma(T)$. We treat the electron as a classical spherical particle that has a classical electron radius (Lorentz radius) of

$$r_e = (4\pi\varepsilon_0)^{-1} e^2 / m_e c^2 , \qquad (13)$$

which has the numerical value $r_e \approx 2.82\times10^{-15}$ m. Using the equation for linear viscous drag for a particle in water, the coefficient of resistive force in Eq. (9) is

$$b = 6\pi r \eta , \qquad (14)$$

where $r$ is the Stokes radius of the particle (which for a spherical particle is the same as the radius of the particle) and $\eta$ is the temperature-dependent fluid viscosity. Therefore, we obtain the damping constant in Eq. (9) for the electron in water as

$$\gamma(T) = b / m_e = 6\pi r_e \eta(T) / m_e \qquad (15)$$

At $T$=20°C, the viscosity of water is $\eta = 1.0003\times10^{-3}\, Pa\cdot s$, so that $\gamma = 5.83\times10^{13}$ rad/s.

*3.4 Computer simulations*

Using the model described above, we calculated ratios of scattering and extinction cross-sections efficiencies of charged to uncharged water droplets with radii 5nm<$a$<1μm, incident frequency $f$ = 94GHz and $T$=20°C. Log-log plot of computer simulations of the ratios of charged ($Q^*_{\text{ext}}$) to uncharged ($Q_{\text{ext}}$) extinction cross-section efficiencies as a function of the size parameter $x$ is presented in Fig. 6. For small droplets (a<100nm), the model predicts increased forward scattering of MMW from charged mist, which is qualitatively consistent with experimental observations in Figs. 2 and 3. Log-log plot of computer simulations of the ratios of charged ($Q^*_{\text{sca}}$) to uncharged ($Q_{\text{sca}}$) scattering cross-section efficiencies as a function of the size parameter $x$ is presented in Fig. 7. The model predicts increased MMW backscattering from charged mist, which is qualitatively consistent with experimental observations in Figs. 4 and 5. Physical implication of these results is that for sub-micron droplets, charging increases MMW scattering and decreases MMW absorption. One can hypothesize that increased scattering is due to scattering from extra surface electrons. Absorption in water droplets is related to dissipation of energy due to vibration of dipoles in response to the oscillatory applied EM field. Thus, one can hypothesize that decrease in absorption due to charging may be caused by jamming of surface dipoles when extra surface charges are present. Note from Eqs. (1) and (2) that $Q_{\text{sca}} \propto x^4$, while $Q_{\text{ext}} \propto x$. For MMW incident on mist microdroplets $x$<<1, so that MMW reflection signal for either charged or neutral mist is expected to be much smaller than the transmission signal for either charged or neutral mist, as shown by the relative changes between Figs. 2 and 4 and Figs. 3 and 5.

In order to gain a better understanding of the model, we plot real and imaginary components of the surface dielectric function $\varepsilon_s$ as a function of the size parameter $x$ for water droplets with radii 5nm<$a$<1μm, frequency $f$ = 94 GHz and $T$=20°C in Fig. 8. Numerical values of the parameters used in the computer simulations are $\varepsilon_v$ =7.69 + $i$ *13.32, $\omega = 5.9\times10^{11}$ rad/s, $\gamma = 5.8\times10^{13}$ rad/s, $\omega_s = 8.4\times10^{13}$ rad/s for $a$ = 5nm droplet, and since from Eq. (12) $\omega_s \propto 1/a$, $\omega_s = 4.2\times10^{11}$ rad/s for $a$ = 1μm droplet. From Eq. (10),

$$\text{Re}\,\varepsilon_s = -\omega_s^2 / (\omega^2 + \gamma^2) \text{ and } \text{Im}\,\varepsilon_s = \omega_s^2 \gamma / (\omega^3 + \omega\gamma^2) . \qquad (16)$$

Since $\omega << \gamma$, we obtain

$$\mathrm{Re}\,\varepsilon_s \approx -\omega_s^2/\gamma^2 \text{ and } \mathrm{Im}\,\varepsilon_s \approx \omega_s^2/(\omega\gamma). \qquad (17)$$

This explains why $\mathrm{Im}\,\varepsilon_s >> \mathrm{Re}\,\varepsilon_s$ for small droplets, but with increasing size $\mathrm{Im}\,\varepsilon_s \to 0$ and $\mathrm{Re}\,\varepsilon_s \to 0$. Note also that for small droplets $\mathrm{Im}\,\varepsilon_s >> \mathrm{Re}\,\varepsilon_v$ and $\mathrm{Im}\,\varepsilon_s >> \mathrm{Im}\,\varepsilon_v$. We label the quantity in brackets in Eqs. (1) and (2) as $\alpha$ having the numerical value $\alpha$ =0.82 +$i$0.15 for uncharged droplets and

$$\alpha^* = [\varepsilon_v + \varepsilon_s - 1]/[\varepsilon_v + \varepsilon_s + 2] \qquad (18)$$

for charged ones. One can show that for small droplets $|\alpha^*|^2 \approx 1$ and $\mathrm{Im}\,\alpha^* << 1$, whereas $|\alpha|^2 = 0.82$ and $\mathrm{Im}\,\alpha = 0.15$. Then, taking the appropriate ratios, one obtains that for small droplets scattering is slightly higher, while absorption is an order of magnitude smaller than the corresponding quantities for neutral droplets, as can be seen in Figs. 3 and 4. With increasing size, the effect of charge is diminished.

The results of our numerical experiments indicate that no significant sensitivity of the scattering and absorption cross-sections to surface charge is observable for droplets larger than approximately $a$=100nm. For droplets smaller than approximately $a$=5nm, our model predicts that no surface charges will be deposited. On the other hand, existing models and experimental observations indicate that most droplets produced by UA mist generators have sizes larger than 1μm [16]. In order to bridge the gap between theory and experiment, in our future work we will develop more accurate model of scattering from droplets and improve the experimental setup to obtain more quantitative information about the neutral and charged mist. Scattering model we have developed assumes a continuum surface charge model. For a continuum, metal-like surface charge, Coulomb repulsion forces may be ignored, and drift of charges in the external field may be approximated as surface current. However, small droplets carry a discrete number of surface charges, where Coulomb repulsion cannot be ignored. Also, our model ignores thermal fluctuations. However, at ambient temperature, energy of thermal fluctuations $k_B T$ exceeds the energy of the driving force of applied electromagnetic field. The model damping constant presented in Section 3.3 is an empirical one. We expect the model validity range to be limited to water droplets at room temperature. In our future work, we plan to utilize molecular dynamics (MD) simulations for first-principles calculations of surface charge interaction with the electromagnetic field [17]. Mobile surface charge damping constant will be calculated with MD using realistic surface potential.

## 4. Conclusion

We have observed increased forward and backscattering of 94GHz millimeter wave (MMW) from charged mist, as compared to MMW scattering from neutral mist. Ambient temperature mist was produced from de-ionized water with an ultrasonic atomizer (UA), and charging was accomplished with a negative ion generator (NIG) placed near the mist. We have attempted to interpret the experimental results using a classical electrodynamics-based model of scattering from diffusion-charged water droplets. Our model predicts increased forward scattering and increased backscattering of MMW from small droplets, which qualitatively agrees with the observed experimental results. However, charging is predicted to affect droplets in the size range below 100nm, while current literature on UA-generated mist suggests that sizes of most droplets are above 1μm. In order to bridge the gap between theory and experiment, in our future work we will develop more accurate model of scattering from droplets and improve the experimental setup to obtain more quantitative information about the neutral and charged mist.

Our investigation of EM scattering from a single charged water aerosol represents the initial attempt at understanding the problem. In order to obtain accurate description of the interaction of surface charges with electromagnetic radiation we will employ molecular dynamics (MD) simulation of charged water droplets [17]. MD allows simulating water surface with realistic surface charge trapping potential. In addition, hopping of charge from one molecule to another can be calculated. Results of computer simulations will outline the conditions of the scattering method applicability, such as the minimum amount of ionizing radiation detectable, as well as suggest the optimum electromagnetic wavelength and intensity. Extension of the model of scattering from a single droplet to that of an ensemble requires understanding of the ratio of droplets that will undergo charging. Polarimetry may provide additional information about the system, since depolarization ratio in multiple scattering may be different for charged and neutral humid air [18].

In our paper, we chose to consider the mechanism of ionization-related change in dielectric polarizability using the hypothesis of charge-related surface dielectric function of water droplets. However, scattering from random medium can change if refractive index distribution function of the medium changes in response to external stimulus. We will develop a model of ionization-related change in refractive index distribution function of the mist, and the effect of such change on scattering of MMW in our future work.

In our future experimental work, we will set up an environmental chamber in which charged clouds can be generated and well instrumented for humidity, charge density, droplet size, etc. Measurements of droplet size and surface charges on droplets are especially important for model validation. We will use laser scattering techniques for droplet size distribution measurement. We will make time correlated measurements of the millimeter wave reflection and transmission signals in accordance with the timings of charge generation and decay.

**Acknowledgements**

This work was performed under the auspices of the National Nuclear Security Administration, Office of Nonproliferation Research and Development, U.S. Department of Energy, under contract number DE-AC02-06CH11357. Computational resources were provided in part by the Argonne National Laboratory’s Center for Nanoscale Materials CNM 951 grant.

**List of Figure Captions**

Fig.1. Experimental setup for 94GHz MMW forward- and backscattering measurements in neutral and charged humid air. Humidity is produced with an ultrasonic atomizer (UA), and charges are emitted form negative ion generator (NIG). Reflection measurements were performed in bistatic geometry, with a receiver antenna, coupled with 20dB preamplifier, placed next to the emitter antenna. (color online).

Fig.2. Measurement of 94GHz MMW forward scattering in neutral and charged flowing (convective and diffusive) mist: (a) Time-dependent change in MMW transmitted power *T*; (b) Time-dependent profile of air ionization.

Fig.3. Measurement of 94GHz MMW forward scattering in neutral and charged stationary (diffusive only) mist: (a) Time-dependent change in MMW transmitted power *T*; (b) Time-dependent profile of air ionization.

Fig.4. Measurement of 94GHz MMW backscattering from neutral and charged flowing (convective and diffusive) mist: (a) Time-dependent change in MMW reflected power *R*; (b) Time-dependent profile of air ionization.

Fig.5. Measurement of 94GHz MMW backscattering from neutral and charged stationary (diffusive only) mist: (a) Time-dependent change in MMW reflected power *R*; (b) Time-dependent profile of air ionization.

Fig.6. Log-log plot of computer simulations of the ratios of charged ($Q^*_{ext}$) to uncharged ($Q_{ext}$) extinction cross-section efficiencies as a function of size parameter $x$ for water droplets with radii 5nm<$a$<1μm, frequency $f$ = 94GHz and $T$=20°C. For small droplets (a<100nm), the model predicts increased MMW forward scattering in ionized mist, which is qualitatively consistent with experimental observations in Figs. 2 and 3.

Fig.7. Log-log plot of computer simulations of the ratios of charged ($Q^*_{sca}$) to uncharged ($Q_{sca}$) scattering cross-section efficiencies as a function of size parameter $x$ for water droplets with radii 5nm<$a$<1μm, frequency $f$ = 94GHz and $T$=20°C. For small droplets (a<100nm), the model predicts increased MMW backscattering in ionized mist, which is qualitatively consistent with experimental observations in Figs. 4 and 5.

Fig.8. Log-linear plot of real and imaginary components of the surface dielectric function $\varepsilon_s$ as a function of the size parameter $x$ of for water droplets with radii 5nm<$a$<1μm, frequency $f$ = 94 GHz and $T$=20°C. (color online).

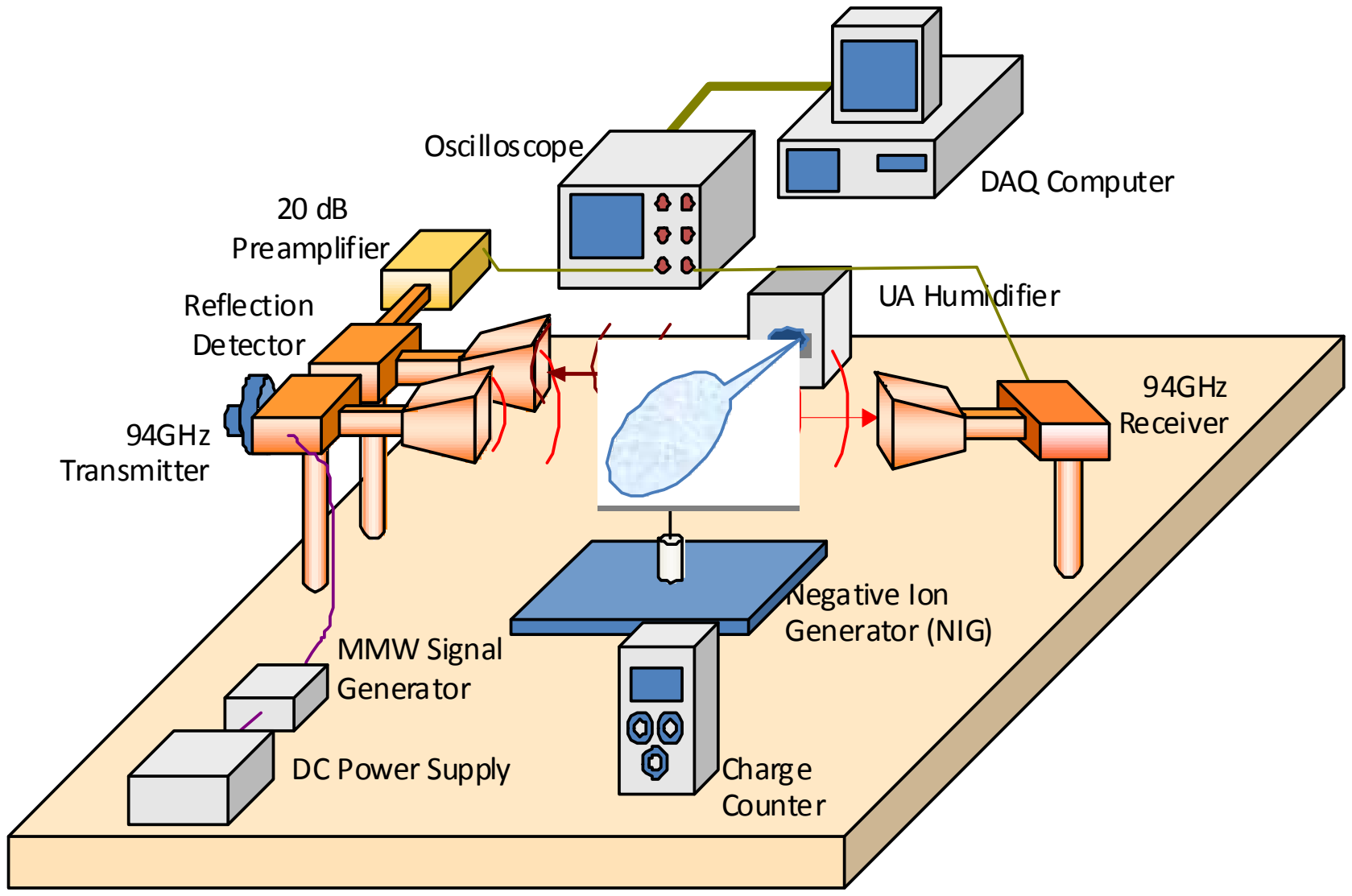

Figure 1

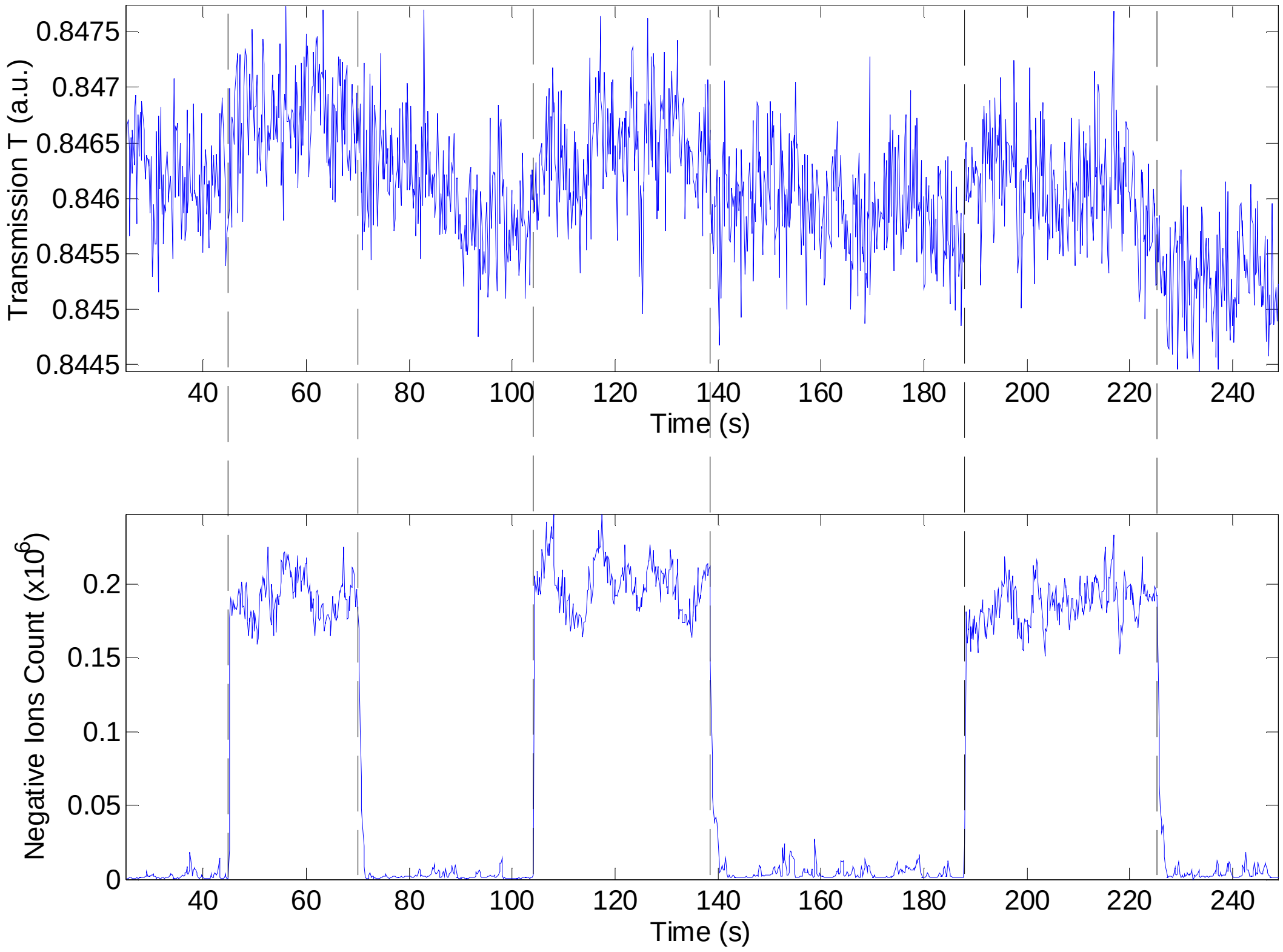

Figure 2

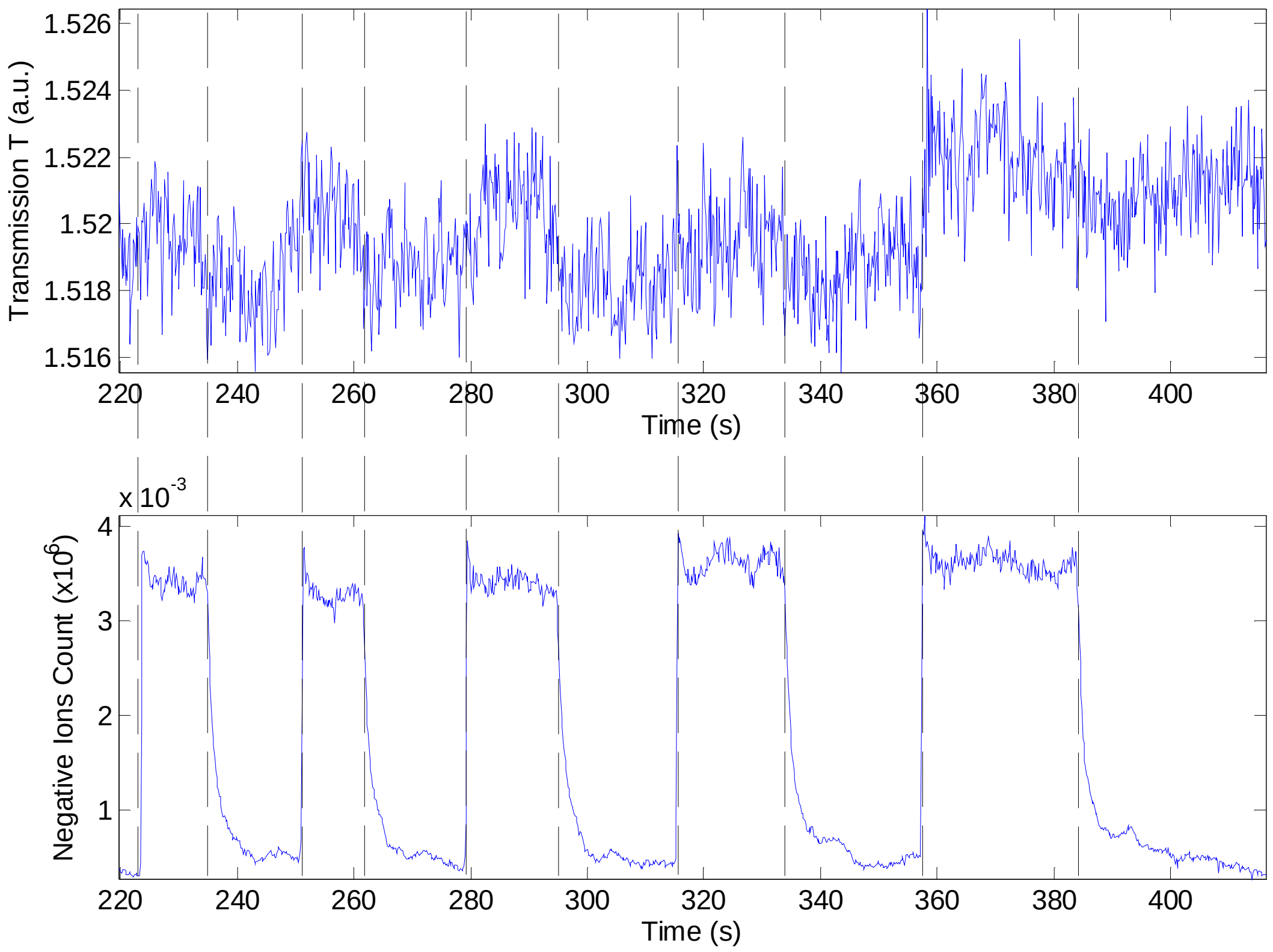

Figure 3

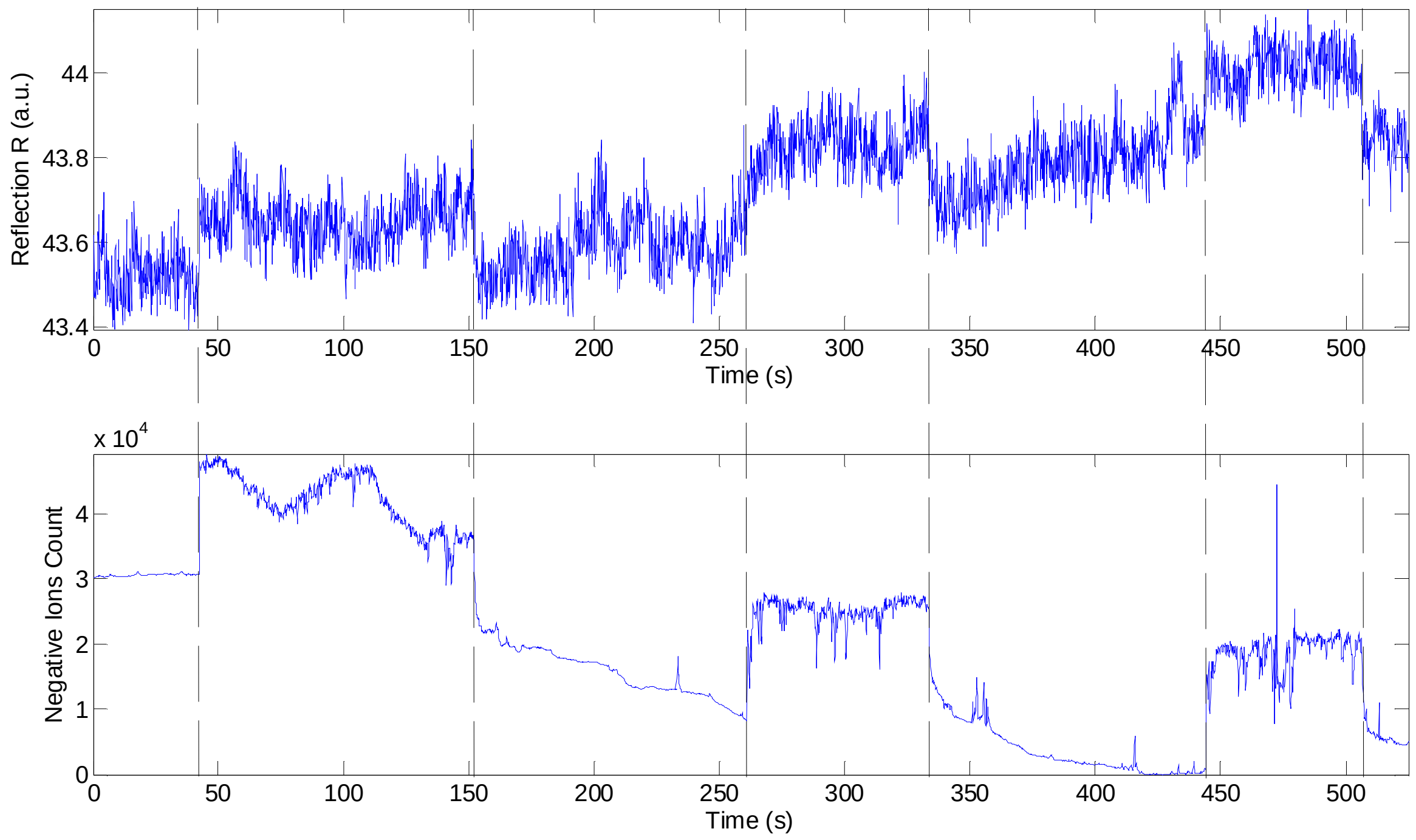

Figure 4

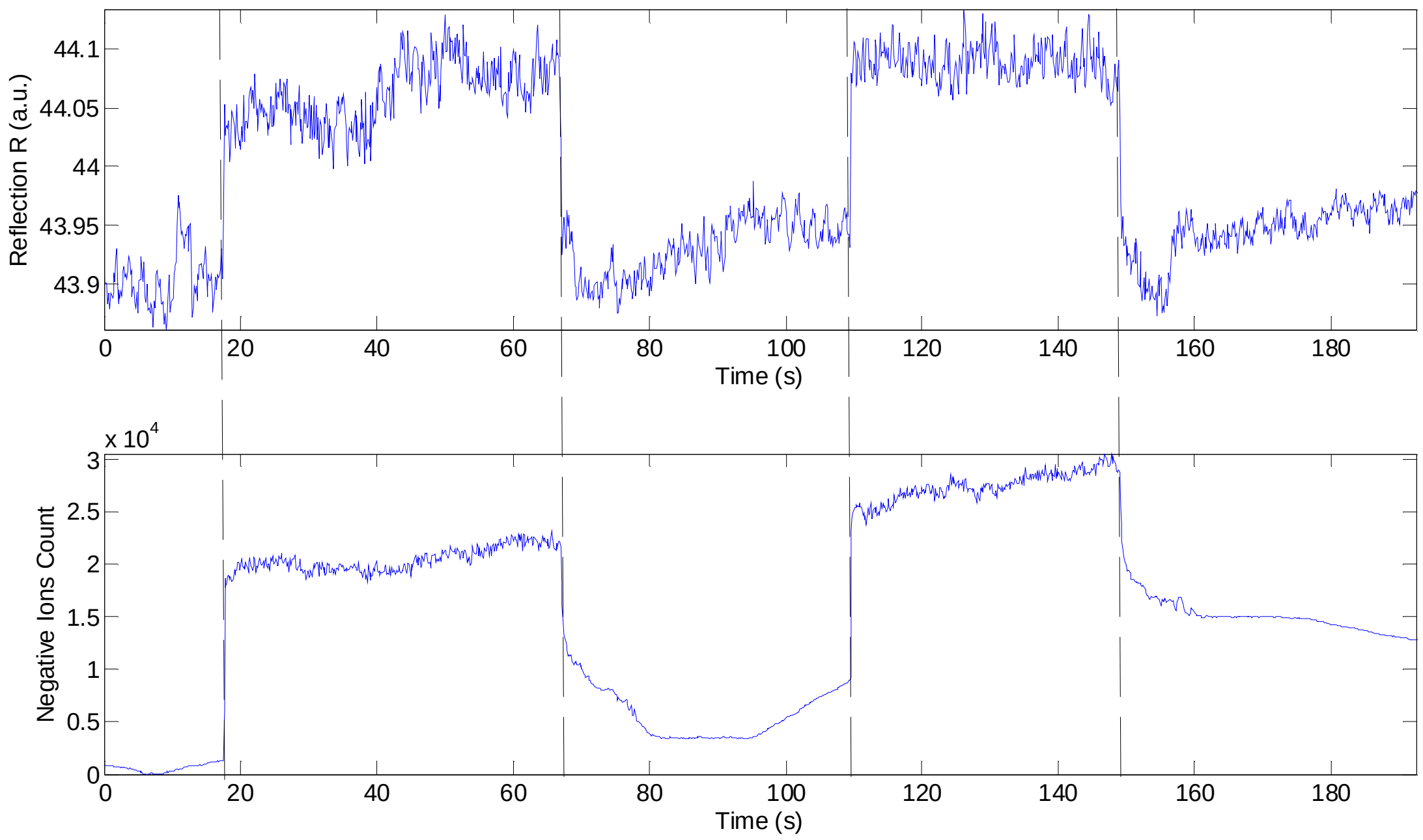

Figure 5

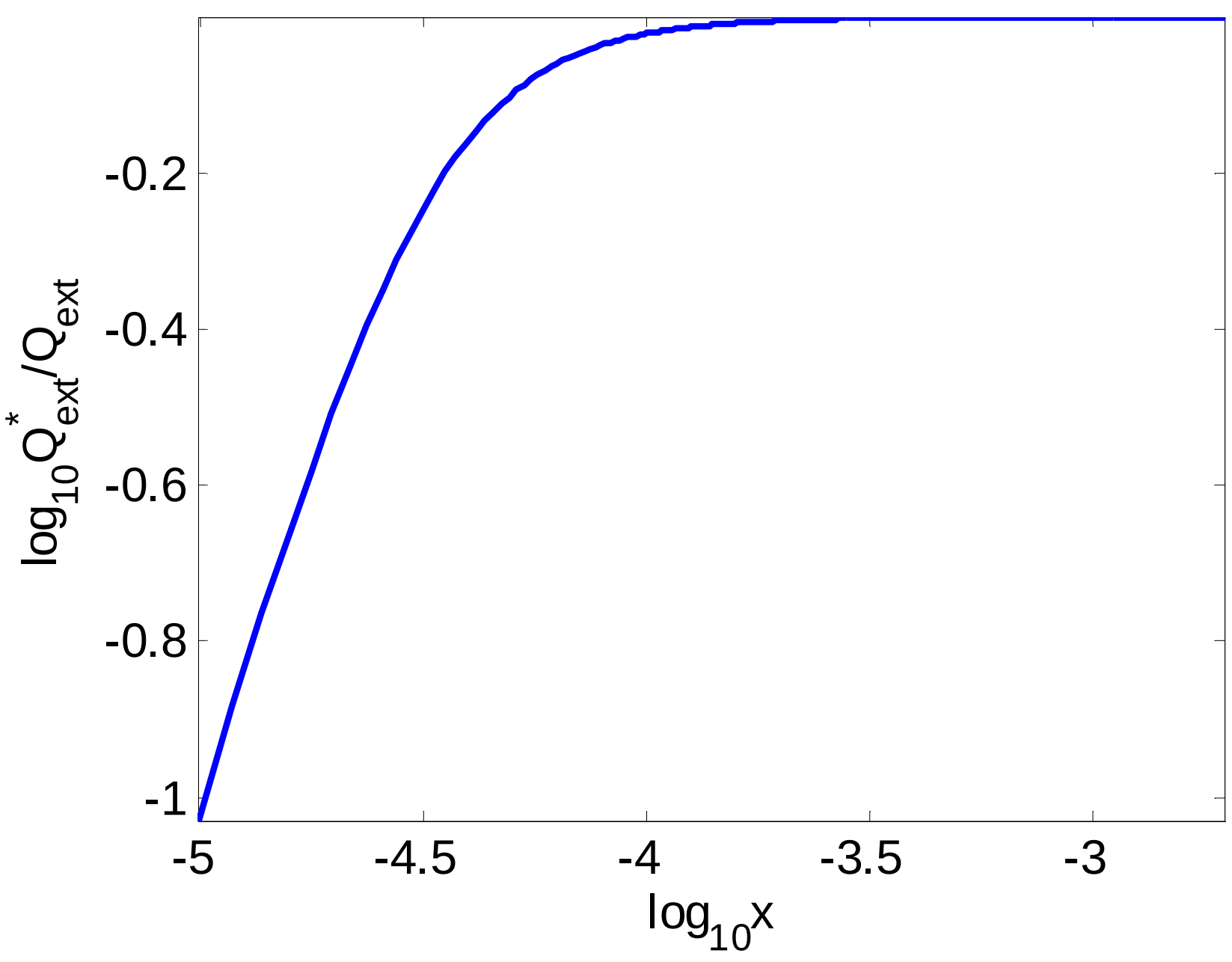

Figure 6

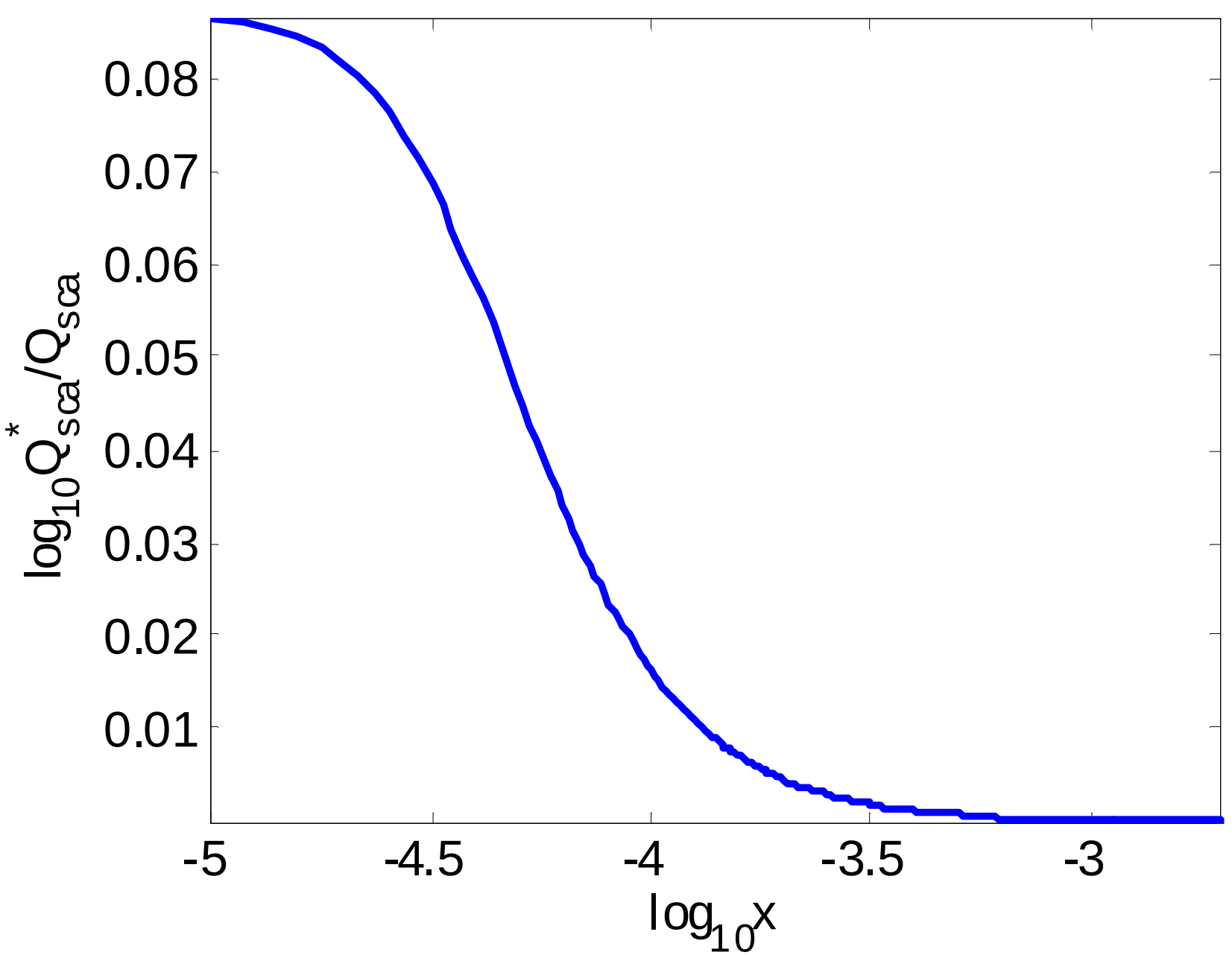

Figure 7

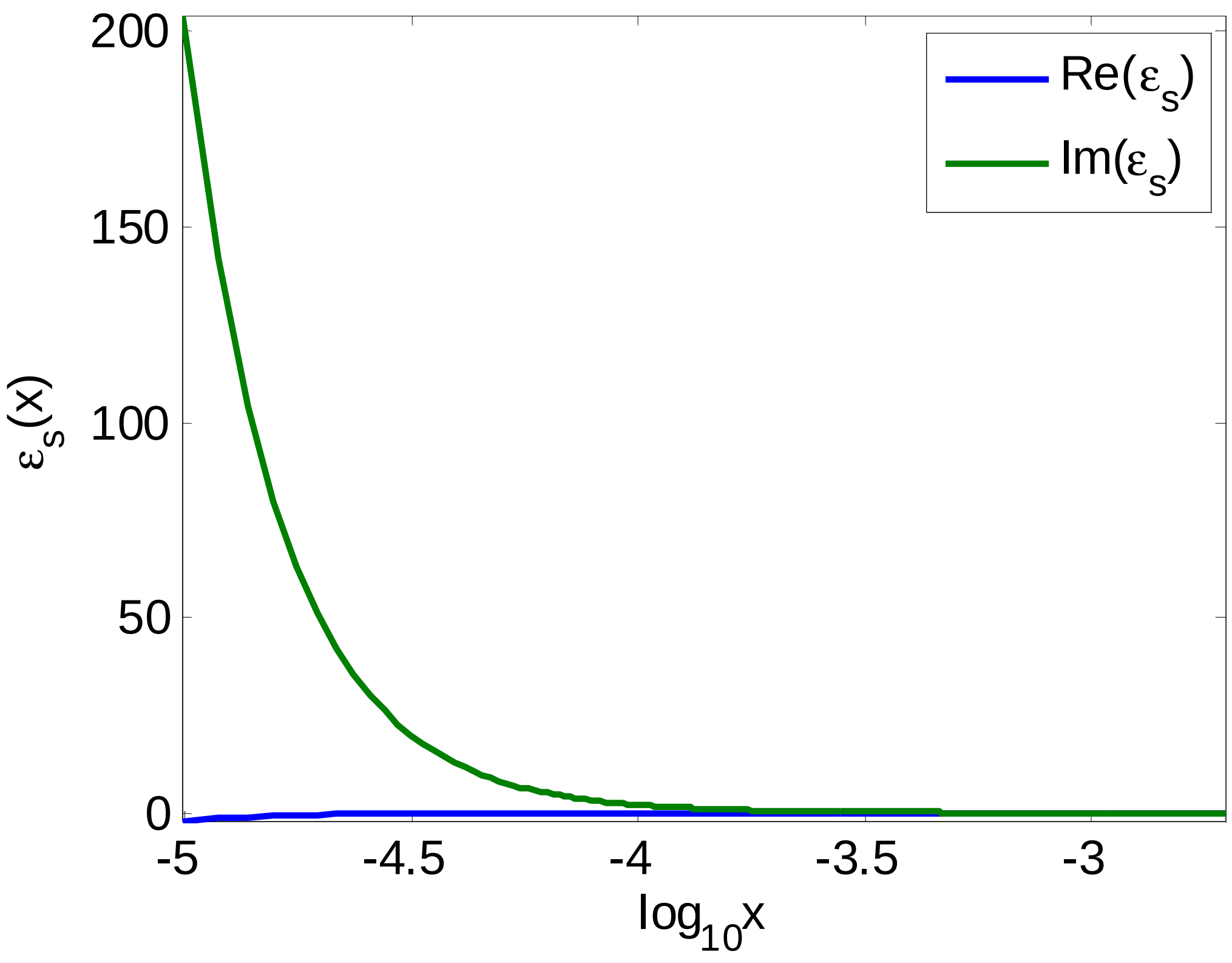

Figure 8